\documentclass[aps,prb,twocolumn,groupedaddress]{revtex4}
\usepackage{graphicx}
\usepackage{amssymb}
\usepackage{amsmath}
\usepackage{bm}
\usepackage{xcolor}

\newcommand{\ket}[1]{|{#1}\rangle}

\newcommand{\abs}[1]{\left|{#1}\right|}

\newcommand{\Hssh}{H_\textrm{SSH}}

\newcommand{\Heff}{H_\mathrm{eff}}

\begin{document}

\title{Chiral symmetry and bulk--boundary correspondence in
  periodically driven one-dimensional systems } 

\author{J.~K.~Asb\'oth} 
\affiliation{Institute for Solid State Physics
  and Optics, Wigner Research Centre, Hungarian Academy of Sciences,
  H-1525 Budapest P.O. Box 49, Hungary} 

\author{B. Tarasinski} \affiliation{Instituut-Lorentz, Universiteit
  Leiden, P.O. Box 9506, 2300 RA Leiden, The Netherlands}

\author{P. Delplace} \affiliation{Laboratoire de Physique, Ecole
  Normale Superieure de Lyon, 47 allee d’Italie, 69007 Lyon, France}

\date{May 2014}
\begin{abstract}

In periodically driven lattice systems, the effective (Floquet)
Hamiltonian can be engineered to be topological: then, the principle
of bulk--boundary correspondence guarantees the existence of robust
edge states.  However, such setups can also host edge states not
predicted by the Floquet Hamiltonian. The exploration of such edge
states, and the corresponding unique bulk topological invariants, has
only recently begun. In this work we calculate these invariants for
chiral symmetric periodically driven one-dimensional systems. We find
simple closed expressions for these invariants, as winding numbers of
blocks of the unitary operator corresponding to a part of the time
evolution. This gives a robust way to tune these invariants using
sublattice shifts. We illustrate our ideas on the periodically driven
Su-Schrieffer-Heeger model, which, as we show, can realize a discrete
time quantum walk: this opens a useful connection between periodically
driven lattice systems and discrete time quantum walks.
Our work helps interpret the results of recent simulations where a
large number of Floquet Majorana fermions in periodically driven
superconductors have been found.

\end{abstract}
\pacs{74.45.+c, 71.10.Pm, 73.23.-b, 74.72.-h}
\maketitle

Controlling the topological phases of matter is an important challenge
in solid state physics. In the recent years, periodic driving has
emerged as an important tool to meet this challenge.  Topologically
protected edge states, the hallmarks of topological phases, have been
predicted and observed in periodically driven systems, such as
materials irradiated by light\cite{floquet_topological,dora_optical,
  floquet_hall_theory,floquet_surface_ti}, in shaken optical
lattices\cite{sengstock_shaken,reichl2014floquet}, and in photonic
crystals\cite{rechtsman_photonic_2013}.  In the above cases, the
principle of bulk--boundary correspondence\cite{schnyder_tenfold} was
applied to the effective (Floquet) Hamiltonian of the periodically
driven system.

The variety of topological phases that periodically driven systems can
display, however, is much wider than those of their Floquet
Hamiltonians, and the systematic exploration of these phases has only
just begun\cite{rudner_driven}.
An important example is the case of periodically driven
one-dimensional topological superconductors,where, the bulk
$\mathbb{Z}_2$ invariant is replaced by a pair of $\mathbb{Z}_2$
invariants, whose calculation necessitates information beyond that
represented by the Floquet
Hamiltonian\cite{akhmerov_majorana_driven}. The edge states then are
the Floquet Majorana fermions, with potential applications in quantum
information processing\cite{floquet_qubits}. Such states, not
predicted by the bulk Floquet Hamiltonian, have also been observed in
optical realization of a one-dimensional quantum
walk\cite{kitagawa_observation}.

Simulations of one-dimensional periodically driven superconductors
have shown that they can host a large number of Floquet Majorana
fermions at their ends\cite{many_majoranas,thakurathi_2013}. This can
be explained by an extra chiral symmetry (CS) of the Floquet
Hamiltonian, which prevents Majorana fermions on the same sublattice
from recombining into complex fermions. Although this explanation is
sufficient in some cases\cite{many_majoranas,thakurathi_2013}, it
cannot be general as it only relies on the Floquet
Hamiltonian. Thus, the question is still open: what are the bulk
topological invariants for periodically driven systems with CS?

In this paper, we find the bulk--boundary correspondence for
periodically driven one-dimensional quantum systems with chiral
symmetry, building on the theory of CS in discrete-time quantum
walks\cite{kitagawa_exploring,kitagawa_introduction,asboth_prb,asboth_2013}.
We show how CS can be ensured in a periodically driven system, whose
time evolution in a period starts with a unitary operator $F$, by
choosing an appropriate second part for the period.  We show that the
topological invariants predicting the number of 0 and $\pi$
quasienergy end states are the winding numbers of the blocks of $F$ in
a canonical basis. Our formulas give a direct recipe to tune the
topological invariants using a sublattice shift operation. We give an
example of how to realize this operation in the simplest periodically
driven one-dimensional Floquet insulator with CS, the periodically
driven Su-Schrieffer-Heeger (PDSSH) model. We show how this model
realizes a discrete-time quantum walk, and how this can be used to
calculate the topological invariants of particle-hole symmetric
quantum walks.
     
\emph{Floquet formalism. }
We consider periodically driven single-particle lattice Hamiltonians,
$H(t+1) = H(t)$.
The long-time dynamics of $H(t)$, i.e., over many periods, is governed
by the time-evolution operator of one period, the Floquet operator $U(\tau)$,
\begin{align}
U(\tau) &= 
\mathbb{T} 
e^{-i \int_\tau^{\tau+1} H(t) dt},
\end{align}
where $\mathbb{T}$ stands for time ordering. 
If at time $\tau$ the system is in an eigenstate $\ket{\Psi}$ of the
Floquet operator, $U(\tau) \ket{\Psi} = e^{-i\varepsilon}\ket{\Psi}$,
then at all times $\tau+n$, for $n\in \mathbb{Z}$, it will be in state
$e^{-i n \varepsilon} \ket{\Psi}$. In this sense, the periodically
driven system acts as a stroboscopic simulator of the effective
(Floquet) Hamiltonian $\Heff$,
\begin{align}
\Heff(\tau) &= i \mathrm{ln} U(\tau).
\end{align}
We fix the branch of the logarithm by restricting the eigenvalues
$\varepsilon$ of $\Heff$, the quasienergies, to $-\pi< \varepsilon \le
\pi$. 

The Floquet operator $U(\tau)$, and thus also the effective
Hamiltonian $\Heff(\tau)$, depend on the choice of the starting time
of the period, $\tau$. Changing $\tau$ amounts to a unitary
transformation of the Floquet operator and the effective Hamiltonian
(quasienergies are independent of $\tau$).

\emph{Chiral symmetry of periodically driven systems. } Ensuring CS of
the periodically driven system amounts to ensuring that there is an
initial time $\tau$ such that the corresponding effective Hamiltonian
has CS, i.e., there is a unitary, Hermitian, and local (within a unit
cell) operator $\Gamma$, that satisfies
\begin{align}
\Gamma \Heff(\tau) \Gamma &= - \Heff(\tau) &\Leftrightarrow
\,\, \Gamma U(\tau) \Gamma &= U^{-1}(\tau).
\end{align}

The effective Hamiltonian does not inherit CS from the instantaneous
Hamiltonian, as is the case with particle--hole
symmetry\cite{kitagawa_periodic}. However, CS of the periodically
driven system is ensured if there is an intermediate time $0 < t_1 <1$
that splits the period into a first and second part in a special
way. Let $F$ denote the time evolution of the first part of the cycle,
\begin{align}
F &= \mathbb{T} e^{-i\int_{\tau'}^{\tau'+t_1}H(t) dt}.
\label{eq:F_first}
\end{align}
The second part of the cycle has to fulfil  
\begin{align}
\Gamma F^\dagger \Gamma &= \mathbb{T} e^{-i\int_{\tau'+t_1}^{\tau'+1}H(t) dt}.
\label{eq:F_second}
\end{align}
It is easy to check that in that case, 
not only $U' \equiv U(\tau')$, but also $U'' \equiv U(\tau'')$ have CS, 
where $\tau'' = \tau'+t_1$. These Floquet operators read 
\begin{align}
U' &= \Gamma F^\dagger \Gamma F;& 
U'' &= F \Gamma F^\dagger \Gamma.
\label{eq:U_from_F}
\end{align}

\emph{Topological invariants of the effective Hamiltonians due to
  chiral symmetry. } Consider a one-dimensional Floquet insulator: a
long chain, with a translation invariant insulating bulk part, whose
quasienergy spectrum has gaps around $\varepsilon=0$ and $\pi$.
If the system has CS, a local basis transformation can be performed
that diagonalizes $\Gamma$, so that each lattice site has a sublattice
index A or B, defined via the projectors $\Pi_{A/B} =
(1\pm\Gamma)/2$. We call such a basis a \emph{canonical basis}. For
the system to be a Floquet insulator, the number of A and B sites in
each bulk unit cell has to be equal (or else the system would have
flat bands at $0$ or $\pi$ quasienergy). We denote this number by
$N$. In a canonical basis, the CS operator acts in each unit cell
independently, as
$\Gamma = \sigma_z \otimes 1_N$.

The spectrum of an effective Hamiltonian with CS is symmetric:
stationary states $\ket{\Psi'}$ of $\Heff'$ with quasienergy
$\varepsilon\neq 0, \pi$ have chiral symmetric partners $\Gamma
\ket{\Psi'}$, that are also eigenstates with quasienergy
$-\varepsilon$. Such states can be chosen to have equal support on
both sublattices. The system can also host states $\ket{\Psi'}_{L/R}$
with quasienergy $\varepsilon=0$ or $\pi$, whose wavefunctions are
expelled from the bulk to the left/right by the gaps in the bulk
spectrum. These \emph{end states} can be
chosen to have support only on one sublattice.

The effective Hamiltonians $\Heff'$ and $\Heff''$ have CS, as per
Eqs.~\eqref{eq:U_from_F}, and thus can be assigned topological
invariants $\nu'$ and $\nu''$.  These are obtained by standard
procedure\cite{schnyder_tenfold}, whereby we first isolate the bulk
part of $\Heff'$ and $\Heff''$, by imposing periodic boundary
conditions on the translation invariant central part of these
Hamiltonians, and taking the thermodynamic limit. The bulk
Hamiltonians are periodic functions of the quasimomentum $k\in
[-\pi,\pi)$, and, in the canonical basis, are block off-diagonal,
\begin{align}
\Heff(k) &=
\begin{pmatrix}
0 & h(k) \\
{h}^\dagger (k) & 0
\end{pmatrix}.
\end{align}
Here, and later on, $\Heff$ refers to either of $\Heff'$ or
$\Heff''$, and similarly for $U$ and $h$. 
The topological
invariants are 
\begin{align}
\nu' &= \nu[h'];& \nu'' &= \nu[h''],
\end{align}
where the function $\nu[h]$ is a winding number,
\begin{align}
\nu[h] &= \frac{1}{2\pi i} \int_{-\pi}^{\pi} dk \frac{d}{dk} \text{ln
} \text{det } h (k).
\label{eq:winding_def}
\end{align}
These integers cannot change under adiabatic deformation of the bulk
Hamiltonians, and so are equal to the winding numbers of the flat band
limits of these Hamiltonians, which are the topological invariants of
Ryu et al \cite{schnyder_tenfold}. They can be interpreted as the
dimensionless bulk sublattice
polarization\cite{mondragon2013topological} of the effective
Hamiltonians, at times $\tau'$ and $\tau''$.

\emph{Topological invariants of the driven system.}  To derive the
topological invariants of the periodically driven system, we start by
adopting the results obtained for discrete-time quantum walks (DTQW)
with CS\cite{asboth_2013} to periodically driven systems. The
derivations follow very closely those of
Ref.~\onlinecite{asboth_2013}, and so we omit them here, but for
completeness, we give details in Appendix \ref{app:dtqw_to_driven}.
As with DTQWs, also in periodically driven systems, the wavefunctions
of quasienergy $\pi$ end states switch sublattices as they evolve from
time $\tau'$ to $\tau''$, and so, neither $\nu'$, nor $\nu''$, on
their own, give useful information about the number of end states
(observations to the contrary in specific
models\cite{many_majoranas,thakurathi_2013} do not generalize). The
winding numbers $\nu'$ and $\nu''$ must be combined to obtain the bulk
topological invariants controlling the number of end states,
\begin{align}
\nu_0 &= \frac{\nu' + \nu''}{2};& \nu_\pi &= \frac{\nu' - \nu''}{2}.
\label{eq:asboth_obuse}
\end{align}

We now proceed to simplify Eqs.~\eqref{eq:asboth_obuse}, and
express them using the blocks of $F$ in the canonical basis:
\begin{align}
F(k) &= \begin{pmatrix}
a(k) & b(k)\\
c(k) & d(k)
\end{pmatrix}.
\label{eq:F_canonical}
\end{align}
Along the way, we will use simple properties of the function
$\nu[A(k)]$ of Eq.~\eqref{eq:winding_def}: $\nu[AB]=\nu[A]+\nu[B]$ and
$\nu[A^\dagger] = -\nu[A]$, for arbitrary $A(k)$ and $B(k)$.

There are two constraints on the winding numbers of the blocks of the
Floquet operator $F$ representing the first part of the drive cycle,
both following from the unitarity of $F$. First, substituting
Eqs.~\eqref{eq:F_canonical} directly into $F(k) F(k)^\dagger = 1$
gives $a c^\dagger = - b d^\dagger$. Taking the winding numbers of the
two sides gives 
\begin{align} 
\nu[c]-\nu[a] &= \nu[d]-\nu[b].
\label{eq:winding_diag}
\end{align} 
Second, $F$ represents an operation on an open chain, terminated at
its ends.  Thus, the average displacement of a state in the bulk, with
this average going over all possible states, has to be zero:
Otherwise, unitarity of $F$ would be violated in the end regions.
This average displacement is given by the winding number of $F$ itself
\cite{kitagawa_periodic}, which, since $F$ is unitary, can be written
as
\begin{align}
\nu[F] &= \frac{1}{2\pi i} \int dk \text{Tr } F^\dagger(k)
\frac{d}{dk} F(k). 
\end{align}
Inserting the decomposition of $F$ in the
canonical basis, Eq.~\eqref{eq:F_canonical}, into $\nu[F]=0$, gives 
\begin{align}
\nu[F] &= \nu[a]+\nu[c]+\nu[b]+\nu[d] = 0.
\label{eq:winding_compatible}
\end{align}


To use the relations derived above, 
we note, that 
\begin{align}
U &= e^{-i \Heff} = \cos \Heff - i \sin \Heff.
\end{align}
Because of the block off-diagonal structure of $\Heff$, 
the first term in the sum above corresponds to the block diagonal and
the second to the block off-diagonal parts of $U$. Now since
$\text{sign}{(\varepsilon)} = \text{sign }{(\sin \varepsilon)}$ for
$\varepsilon \in [-\pi,\pi]$, the winding number of $\Heff$ is the
same as that of $\sin \Heff$.  Therefore, in
Eq.~\eqref{eq:winding_def} above, we can substitute the off-diagonal
block of $U$ in a canonical basis: $h \to i U_{12} $. For the
topological invariants of the effective Hamiltonians $\Heff'$ and
$\Heff''$, using Eqs.~\eqref{eq:U_from_F}, substituting the blocks of
$F$, we obtain $\nu' = \nu[a^\dagger b - c^\dagger d]$ and $\nu'' =
\nu[-a c^\dagger + b d^\dagger]$. We can simplify these using the
unitarity of $F$, whereby $a^\dagger b + c^\dagger d=0$ and $a
c^\dagger + b d^\dagger=0$, and the fact that $\nu[\alpha c]=\nu[c]$
for any $\alpha\in\mathbb{C}$. We obtain
\begin{subequations}
\begin{align}
\nu' &= \nu[b] - \nu[a] = \nu[d] - \nu[c]; 
\label{eq:nup_final}\\
\nu'' &= \nu[a] - \nu[c] = \nu[b] - \nu[d]
\label{eq:nupp_final}.
\end{align}
\end{subequations}
Inserting these equations into Eqs.~\eqref{eq:asboth_obuse}, together
with Eqs.~\eqref{eq:winding_compatible} and \eqref{eq:winding_diag},
gives us
\begin{align}
\nu_0 &= \nu[b];& \nu_\pi &= \nu[d].
\label{eq:asboth_tarasinski}
\end{align}
These equations are the central result of our paper: In one-dimensional
periodically driven systems with CS, the windings of the determinant
of the off-diagonal and the diagonal blocks of the Floquet operator in
a canonical basis fix the number of end states at quasienergy 0 and
$\pi$, respectively.

Eqs.~\eqref{eq:asboth_tarasinski} determine the topological invariant
$\nu_0$ ($\nu_\pi$) even if the gap of $\Heff$ at quasienergy
$\varepsilon=\pi$ ($\varepsilon=0$) is closed, a problem raised by
Tong et al.\cite{many_majoranas}.  Consider
\begin{align}
\cos \Heff' & = 
1-2\begin{pmatrix} 
c^\dagger c & 0 \\
0 & b^\dagger b 
\end{pmatrix} 
=
2\begin{pmatrix} 
a^\dagger a & 0 \\
0 & d^\dagger d
\end{pmatrix}-1.
\label{eq:gap_closings}
\end{align}
If there is a quasimomentum $k$ where the gap of $\Heff'$ closes
around $\varepsilon=0$, then $\cos\Heff'(k)$ has a doubly degenerate
eigenvalue $+1$. At that $k$, using the first relation of
Eq.~\eqref{eq:gap_closings}, either $c(k)$ or $b(k)$ (or both) have an
eigenvalue zero. This means $\nu_0$ is not well defined, and neither
are $\nu'$ or $\nu''$. However, $\nu_\pi$ of
Eq.~\eqref{eq:asboth_tarasinski} is still well defined. Similarly, if
at some $k$ the gap of $\Heff'$ around $\varepsilon=\pi$ closes, then,
using the second relation of Eq.~\eqref{eq:gap_closings}, $a(k)$ or
$d(k)$ must have an eigenvalue zero, and $\nu_\pi$ is not well
defined, but $\nu_0$ is.

\emph{Geometrical picture.} In case of a two-band 1D Floquet insulator
with CS, we can give a geometrical interpretation for the topological
invariants $\nu_0$ and $\nu_\pi$.  We relegate details to Appendix
\ref{sec:geometrical_picture}, and just summarize the results here.

Disregarding an irrelevant global phase, the evolution operator for
the first half of the period reads $F(k) = e^{-i
  \vec{f}(k)\vec\sigma}$,
with $\vec{f}(k)$ a three-dimensional real vector inside a unit sphere
of radius $\pi$, all points on whose surface are identified with each
other, and $\vec{\sigma}$ the vector of Pauli matrices.  As $k$
traverses the Brillouin zone $[ -\pi,\pi[ $, $\vec{f}(k)$ describes a
    directed, smooth, closed loop.  If the gap around $\varepsilon=0$
    is open, the loop of $\vec{f}(k)$ cannot touch the $z$-axis or the
    surface of the sphere, and we find that the invariant $\nu_0$ is
    given by the winding of the loop around the $z$ axis. If the gap
    of $\Heff$ around $\varepsilon=\pi$ is open, the path of
    $\vec{f}(k)$ cannot touch the circle in the $xy$ plane of radius
    $\pi/2$. In that case, $\nu_\pi$ is given by the winding of the
    loop around that circle.

\emph{Tuning the invariants.}
Formulas \eqref{eq:asboth_tarasinski} allow for a simple way to tune
the topological invariants of a periodically driven system, using a
unitary sublattice shift operation $S(n)$, whose bulk part reads 
\begin{align}
S(n,k) &= \text{exp}(-i n \Gamma k). 
\label{eq:def_S}
\end{align}
In the bulk, $S(n)$ displaces sites on sublattice $A$ ($B$) to the
right (left) by $n$ sites. Therefore, at the left/right end, under the
effect of $S(n)$, $n$ states must switch sublattices, transitioning
$B\to A$ / $A \to B$ (if $n$ is negative, vice versa). How this
transition happens depends on the details of $S(n)$ that have no
influence on the topological invariants (nor on the
number of end states).

To tune the invariants of a periodically driven system, obeying 
Eq.~\eqref{eq:U_from_F}, with some $F = F^{(0)}$, insert extra
sublattice shifts before and after $F^{(0)}$, 
\begin{align}
F^{(1)} &= S(m) F^{(0)} S(n).
\label{eq:new_edge_states}
\end{align}
Substituting into  Eqs.~\eqref{eq:asboth_tarasinski}, we obtain
directly the topological invariants of the modified driven system,
\begin{align}
\nu^{(1)}_0 &= \nu_0^{(0)} + m-n;& \nu^{(1)}_\pi &= \nu_\pi^{(0)} -m-n.
\end{align}

\emph{Example: the periodically driven SSH model. }
We now illustrate the concepts introduced above on the PDSSH model,
given by
\begin{align} 
\Hssh(t) &=\sum_{j=1}^{M} \left( 
v(t) c_{2j} c_{2j-1}^\dagger + w(t) c_{2j+1}
c_{2j}^\dagger \right) + \mathrm{h.c.},  
\label{eq:H_ssh}
\end{align} 
where $c_{x}$ annihilates the fermion on site $x$. For simplicity, we
keep the intracell hopping amplitudes $v(t)$ and the intercell hopping
amplitudes $w(t)$ real, homogeneous in space, and modulated
periodically, with period 1.  We fix open boundary conditions by
identifying $c_{2M+1}=0$ (as opposed to periodic boundary conditions,
which would require $c_{2M+1}=c_1$).

The sublattice shift operator $S(n)$ can be
realized\cite{rudner_driven} by the following drive sequence: a pulse
of $v$ of area $\pi/2$, followed by a pulse of $w$ of area $-\pi/2$.
This allows us to realize a discrete time quantum walk as a
periodically driven lattice Hamiltonian.

\begin{figure}[] 
\includegraphics[width=0.5\linewidth]{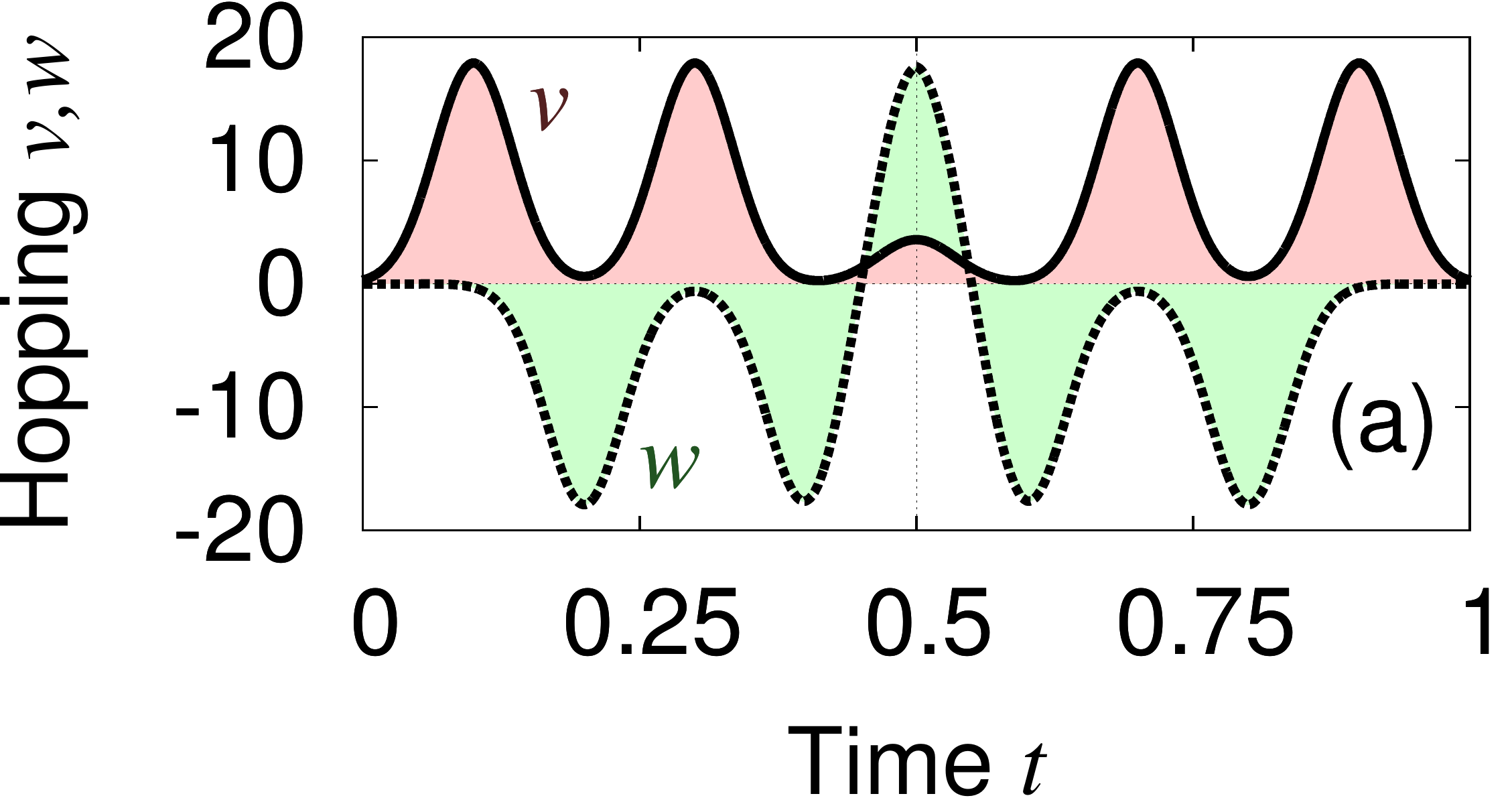}
\hspace{0.3cm}
\includegraphics[width=0.3\linewidth]{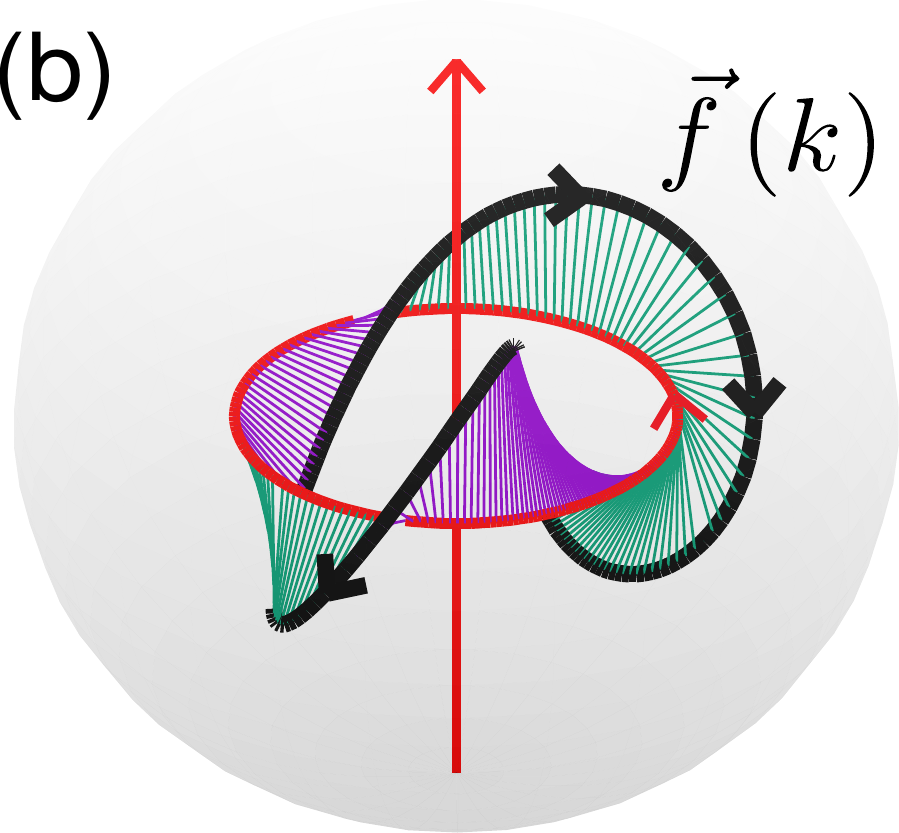}
\hspace{0.5cm}
\includegraphics[width=\linewidth]{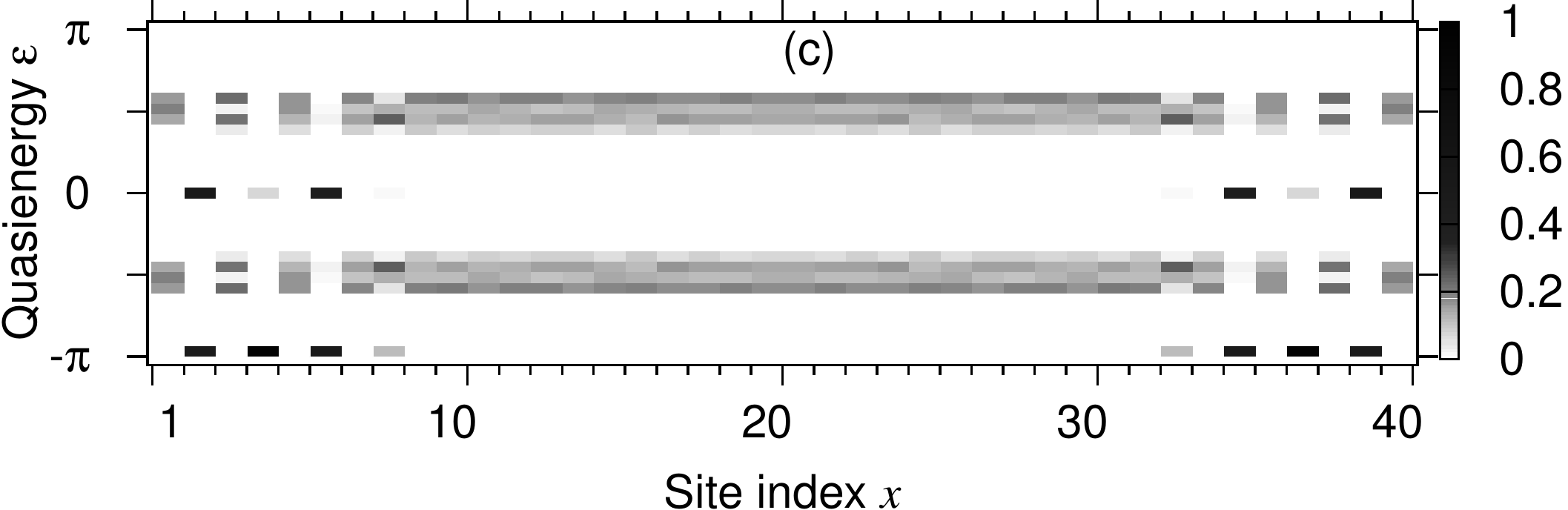}
\includegraphics[width=\linewidth]{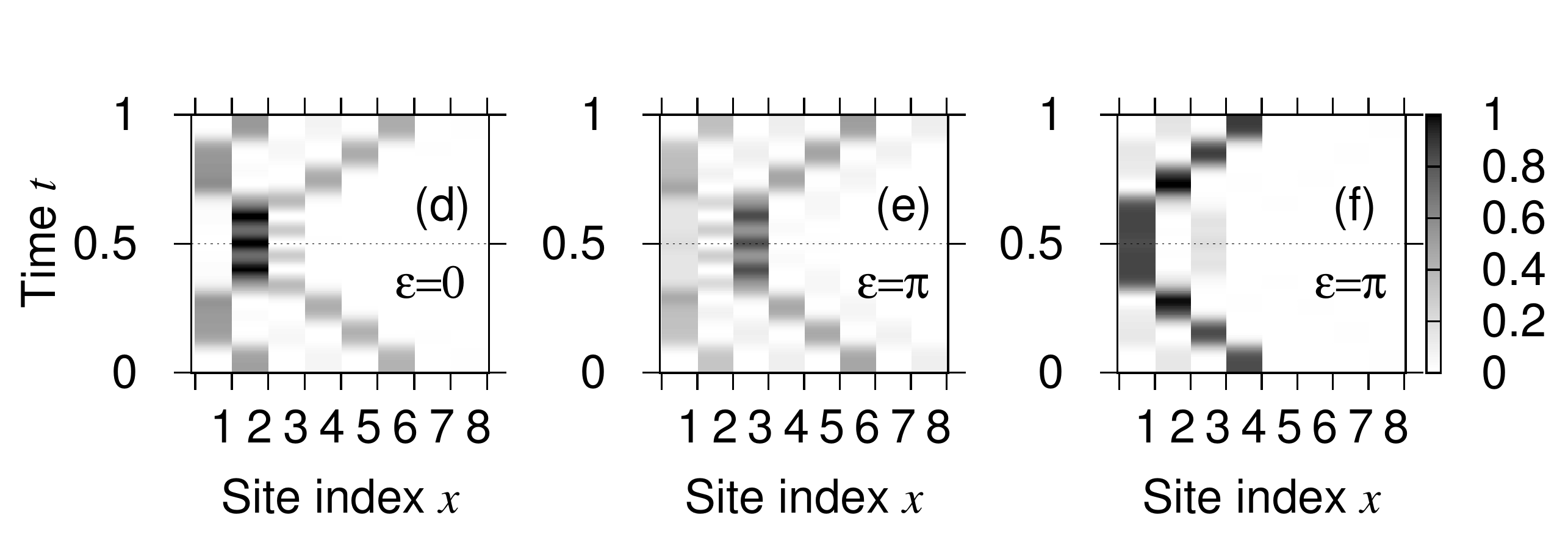}
\caption{Floquet eigenstates of a periodically driven SSH chain of 40
  sites. (a) Time dependence of the intracell (continuous) and
  intercell (dotted) hopping amplitudes. (b) The curve $\vec{f}(k)$,
  which winds -1 times around the z axis (red) and -2 times around the
  circle of radius $\pi/2$ on the $xy$ plane, showing that 
  $\nu_0=-1$ and $\nu_\pi=-2$.  (c) Local Density of States of the
  effective Hamiltonian $\Heff(0)$.
  (d) Time evolution of the position distribution $\abs{\langle
    \Psi(t) | x \rangle}^2$ of the single end state with
  $\varepsilon=0$, and (e,f) of two orthogonal end states with
  $\varepsilon=\pi$. 
\label{fig:time_wavefunctions}
}
\end{figure}

As a concrete example, we consider the PDSSH model on an open chain of
40 sites ($M=20$ unit cells).  The drive sequence, shown in
Fig.~\ref{fig:time_wavefunctions}(a), consists of a train of nine
pulses, chosen to be Gaussian for numerical convenience, applied to
$v$ and $w$ homogeneously.  We ensure CS by way of
Eq.~\eqref{eq:U_from_F}, with $t_1=0.5$, by choosing both $v(t)$ and
$w(t)$ to be even functions of time.

We follow the recipe of Eq.~\eqref{eq:new_edge_states}, to realize
$\nu_0=-1, \nu_\pi=-2$.  The role of role of $F^{(0)}$ is played by
the first half of the central Gaussian pulse, where $w=5v$: thus, it
is a short pulse $e^{-i\pi/2 H_{1}}$, where $H_1$ is an SSH
Hamiltonian in the topologically nontrivial phase. So, we have
$\nu_0^{(0)}=1, \nu_\pi^{(0)}=0$.  To test the robustness of the
recipe, we realize the sublattice displacement $S(n=2)$ only
approximately by allowing considerable overlaps between the $\pi/2$
area pulses of $v$ and the $-\pi/2$ area pulses of $w$.

We find that the bulk topological invariants and the end states agree
perfectly with the theory above. The invariants are the winding
numbers of the curve of Fig.~\ref{fig:time_wavefunctions} (b), which
are $\nu_0=-1, \nu_\pi =-2$. Correspondingly, in the local density of
states, Fig.~\ref{fig:time_wavefunctions} (c), at each end, we find 2
end states at $\varepsilon=\pi$, and 1 end state at $\varepsilon=0$,
exclusively localized on B/A sublattice at the left/right end. The
time dependence of these end states, Fig.~\ref{fig:time_wavefunctions}
(d-f), shows that that they indeed spread over both sublattices at
intermediate times, but return to a single sublattice at $t=0.5$. For
the $0$/$\pi$ energy end states, this is the same/opposite sublattice
as that occupied at $t=0$.

Since we restricted the hopping amplitudes $v$ and $w$ to be real, the
instantaneous SSH Hamiltonian, Eq.~\eqref{eq:H_ssh}, has particle-hole
symmetry (PHS), represented by $\Gamma K$, where $K$ denotes complex
conjugation.
The PDSSH model inherits this symmetry, and therefore, its the end
states are analogous to $0$ and $\pi$ quasienergy Floquet Majorana
fermions. If CS is violated, but PHS is maintained, only the parity of
the number of the Floquet Majorana fermions at each edge and at each
quasienergy $0$,$\pi$ is protected. There is a corresponding pair of
bulk $\mathbb{Z}_2$ topological
invariants\cite{akhmerov_majorana_driven}. In the case of the PDSSH
model, we can follow the construction of Jiang et
al.~\cite{akhmerov_majorana_driven}, and find that the $\mathbb{Z}_2$
invariants can simply be obtained from the complete areas of the
pulses of $v$ and $w$. For details, see Appendix \ref{app:z2}.

\emph{Outlook.}  
The topologically protected states our theory predicts should have
experimental signatures in different kinds of setups. Optical
experiments, where edge states are routinely imaged
directly\cite{kitagawa_observation,PhysRevLett.109.106402}, are in the
best position to test our predictions. Alternatively, in transport
measurements, the end states should give rise to transmission
resonances, similar to the ones predicted for Floquet Majorana
fermions\cite{transport_majorana_floquet}.

Our work leaves a couple of theoretical questions open. 
First, is the decomposition of the drive cycle $U$ into $F$ and
$\Gamma F^\dagger \Gamma$, as per
Eqs.~(\ref{eq:F_first}-\ref{eq:U_from_F}), a necessary requirement for
a periodically driven Hamiltonian to have CS? For previously studied
cases\cite{many_majoranas,thakurathi_2013} we can find such a
decomposition, but if a counterexample were to be found, the theory we
presented here would need to be expanded.
Second, the bulk effective Hamiltonian $\Heff(\tau,k)$ of a
one-dimensional Floquet insulator (with or without CS) is periodic in
both $\tau$ and $k$, and thus has a Chern number. In all the examples
we examined numerically, we found this Chern number to be zero, but
can it take on a nonzero value? If so, what is the physical
interpretation of this number? Last, how can the topological
invariants we found here be formulated in the frequency
domain\cite{rudner_driven}? This is especially an interesting
question, as previous work on the PDSSH model using this
approach\cite{PhysRevLett.110.200403} has not detected the pair of
topological invariants we found. 


We thank J.~Dahlhaus, J.~Li, A.~G\'abris and J.~Edge, for useful
discussions. PD acknowledges useful comments from A. Bernevig. This
research was realized in the frames of TAMOP 4.2.4. A/1-11-1-2012-0001
''National Excellence Program -- Elaborating and operating an inland
student and researcher personal support system'', subsidized by the
European Union and co-financed by the European Social Fund.  This work
was also supported by the Hungarian National Office for Research and
Technology under the contract ERC\_HU\_09 OPTOMECH and the Hungarian
Academy of Sciences (Lend\"ulet Program, LP2011-016). This research
was supported by the Foundation for Fundamental Research on Matter
(FOM) and the Netherlands Organization for Scientific Research
(NWO/OCW).

\bibliography{../walkbib}
\bibliographystyle{apsrev}

\appendix

\section{Derivation of Eqs.~\eqref{eq:asboth_obuse}}
\label{app:dtqw_to_driven}

To derive Eqs.~\eqref{eq:asboth_obuse}, we follow closely the line of
thought of Ref.~\onlinecite{asboth_2013}. We consider an open,
periodically driven chain with CS, which has one bulk and two
ends. Let $n_{A/B,0/\pi}'$ denote the number of end states at the left
end on the $A/B$ sublattice at quasienergy $0/\pi$ of the Hamiltonian
$\Heff'$, and $n_{A/B,0/\pi}''$ the corresponding quantities for
$\Heff''$. The bulk--boundary correspondance for the effective
Hamiltonians $\Heff'$ and $\Heff''$ reads
\begin{subequations}
\begin{align}
\label{eq:edge_walk_p}
\nu' &= n_{A,0}' - n_{B,0}' + n_{A,\pi}' - n_{B,\pi}';\\
\label{eq:edge_walk_pp}
\nu'' &= n_{A,0}'' - n_{B,0}'' + n_{A,\pi}'' - n_{B,\pi}''.
\end{align}
\label{eq:edge_walk}
\end{subequations}

Topologically protected end states of periodically driven
one-dimensional lattices with CS can be divided to two classes: a),
they have quasienergy $0$ and are on the same sublattice at $\tau'$
and $\tau''$, or b) have quasienergy $\pi$ and are on opposite
sublattices. 
Indeed, consider a topologically
protected end state $\ket{\Psi'}$, which is an eigenstate of $U'$ with
eigenvalue $e^{-i\varepsilon}$, with $\varepsilon \in \{0,\pi\}$. It
is only on a single sublattice: $\Gamma \ket{\Psi'} = e^{-i\gamma}
\ket{\Psi'}$, with $\gamma=0/\pi$ corresponding to sublattice $A/B$.
Now consider the same end state at the other special time $\tau''$,
$\ket{\Psi''}=F \ket{\Psi'}$. This is an eigenstate of $U''$ with the
same quasienergy $\varepsilon$.  This state is also on one sublattice
only, because
$
\Gamma F \ket{\Psi'} = \Gamma F \Gamma e^{i\gamma} \ket{\Psi'} = 
\Gamma F \Gamma e^{i(\gamma -\varepsilon)} \Gamma F^{-1} \Gamma F \Psi'
= e^{i(\gamma-\varepsilon)} F \ket{\Psi'}.  
$
So $\ket{\Psi''}$ is on the same (opposite) sublattice as $\ket{\Psi'}$
if $\varepsilon=0$ ($\varepsilon=\pi$). 
This can be written succintly as 
\begin{subequations}
\begin{align}
n_{A,\pi}'' - n_{B,\pi}' &= n_{B,\pi}'' - n_{A,\pi}' = 0;\\
n_{A,0}'' - n_{A,0}' &= n_{B,0}'' - n_{B,0}' = 0.
\end{align}
\label{eq:n_AB}
\end{subequations}

Using Eqs.~\eqref{eq:n_AB} to simplify $\nu'+\nu''$ and $\nu'-\nu''$ from 
Eqs.~\eqref{eq:edge_walk}, we obtain 
\begin{align}
\nu_0 &= \frac{\nu' + \nu''}{2};& \nu_\pi &= \frac{\nu' - \nu''}{2},
\end{align}
which are Eqs.~\eqref{eq:asboth_obuse} we set out to demonstrate.

\section{Geometrical picture}
\label{sec:geometrical_picture}

For a two-band 1D Floquet insulator with CS, we can give a
direct geometrical picture for the topological invariants $\nu_0$ and
$\nu_\pi$.  Since the global phase cannot wind ($F$ cannot have
quasienergy winding), it can safely be disregarded, and the evolution
operator for the first half of the period then reads $F(k) = e^{-i
\vec{f}(k)\vec\sigma}$.
Here $\vec{f}$ is a 3-dimensional vector, of magnitude $f \in
[0,\pi]$
and $\vec{\sigma}$ the vector of
Pauli matrices. The $k$-dependent vector $\vec{f}(k)$ is
restricted inside a spherical ball of radius $\pi$, with all points on the
surface identified with each other.  The $a,b,c,d$ in
Eq.~\eqref{eq:F_canonical} are just complex number valued functions of
$k$,
\begin{align}
F &= \begin{pmatrix}
\cos f -i \sin f \cos \theta & -i \sin f \sin \theta e^{-i\phi} \\
-i \sin f \sin \theta e^{i\phi} & 
\cos f  + i \sin f \cos \theta 
\end{pmatrix},
\label{eq:F_ssh}
\end{align} 
using spherical coordinates.  As $k$ traverses the Brillouin zone,
$\vec{f}(k)$ describes a directed, smooth, closed loop, that can at
some $k$ exit the ball at a point on the surface and reenter at the
same $k$ at the antipodal point.
 
%

If the gap around $\varepsilon=0$ is open, the loop of $\vec{f}(k)$
cannot touch the $z$-axis, nor the surface of the sphere.
Thus, the loop has a well defined winding number around the $z$ axis,
\begin{align}
\nu_0 &= \frac{1}{2\pi} \int dk \frac{d}{dk}
\phi(k).
\end{align}
Since both $f(k),\theta(k) \in ]0, \pi[$ for all $k$, this is the same as 
the winding number $\nu_0$ obtained by substituting \eqref{eq:F_ssh}
into Eq.~\eqref{eq:asboth_tarasinski}. 
  
The gap of $\Heff$ around $\varepsilon=\pi$ closes when $\vec{f}(k)$ is
on the circle on the $n_z=0$ plane of radius $\pi/2$ ($n_z=0$ and
$f=\pi/2$). Thus, if the gap around $\varepsilon=\pi$ is open, the
loop of $\vec{f}(k)$ has a well defined winding number around that
circle.  To calculate this winding number, first discard the $\phi$
information, by setting $\phi=0$. This transforms the 3D closed path
of $\vec{f}(k)$ into a 2D path in a semicircle, with the points on the
circular boundary with the same $x$ coordinate identified. We need the
winding of this path around the single point, $f=\pi/2, n_z=0$. This
is found by deforming the semicircle yet again, by the transformation
$(f\sin \theta, f \cos \theta) \to (\cos f, \sin f \cos \theta) $,
into circle, into whose origin the point $f=\pi/2, n_z=0$ is
mapped. The winding number is then
\begin{align}
\nu_\pi &= \frac{1}{2\pi} \int dk \frac{d}{dk}
\text{arctan} \frac{\cos f(k)}{\sin f(k) \cos \theta(k)}
\end{align}
which is the same as $\nu_\pi$
obtained by substituting Eq.~\eqref{eq:F_ssh}
into Eq.~(\ref{eq:asboth_tarasinski}b).

\section{The $\mathbb{Z}_2 \times \mathbb{Z}_2 $ invariant}
\label{app:z2}

The PDSSH model, Eq.~\eqref{eq:H_ssh}, has particle--hole symmetry
(PHS), represented by $\Gamma K$, where $K$ stands for complex
conjugation. This antiunitary symmetry is inherited by the effective
Hamiltonian from the instantaneous
Hamiltonian\cite{kitagawa_periodic}. 

\begin{figure}[] 
\includegraphics[width=\linewidth]{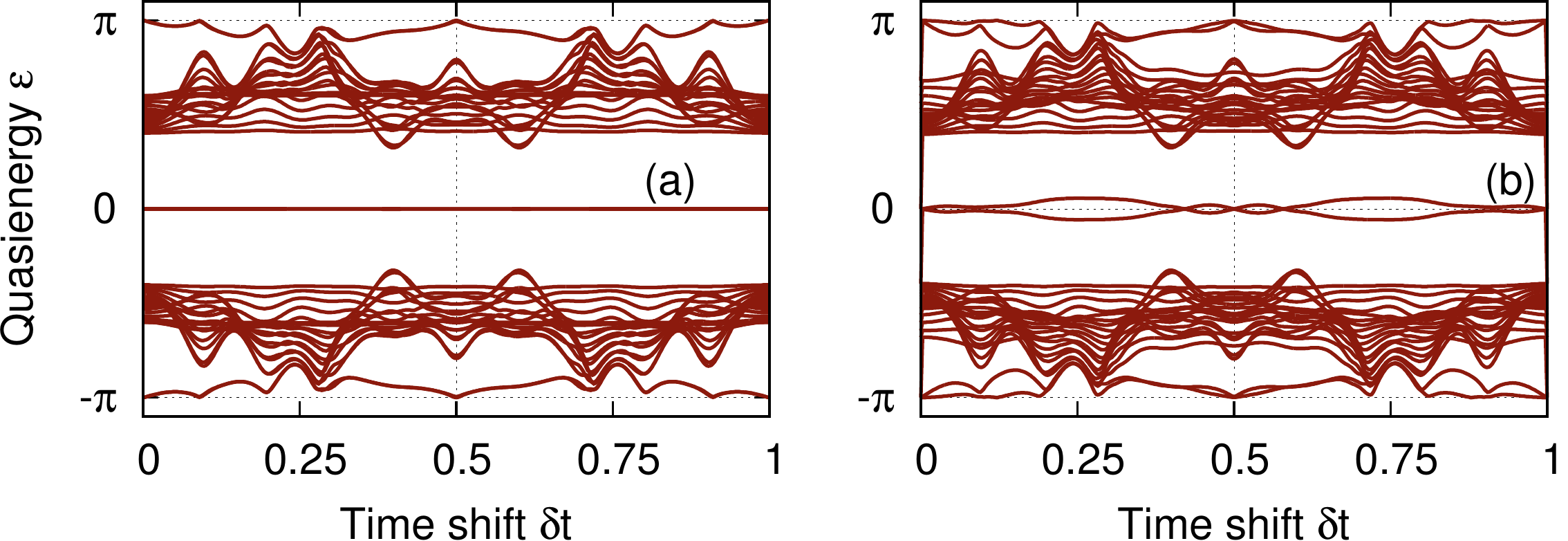}
\caption{Effect of breaking CS by time-shifting the pulse of the
  intracell hopping $v(t)$ with respect to the other pulses.  (a) In
  the PDSSH model, the extra PHS protects the end states at
  $\varepsilon=0$. (b) In the PDRM model, there is no PHS, and all end
  state energies are affected by the time shift.
\label{fig:break_cs}
}
\end{figure}

If we break CS in the PDSSH model, an end state can remain protected
if it can have no PHS partner. This happens whenever the number of end
states at a given energy and at a given end is odd: then, after
breaking CS, a single end state is still protected by PHS.  We
illustrate this on the PDSSH model. If we break CS by delaying the
intracell hopping amplitude $v$ by $\delta t$ with respect to the
intercell hopping $w$ pulses, as shown in Fig.~\ref{fig:break_cs}(a),
the lone end state at $\varepsilon=0$ is still topologically
protected, while the pair of end states at $\varepsilon=\pi$ hybridize
and move away from the edge of the energy Brillouin zone (except for a
time shift of $0.5$, where the conditions for CS are again
fulfilled). To break PHS, we can add a sublattice potential to the SSH
model, obtaining the periodically driven Rice-Mele (PDRM) model,
\begin{align}
H_\text{RM}(t) &= H_\text{SSH}(t)+ u(t) \sum_{x=1}^M \left(c_{2x-1}^\dagger c_{2x-1} -
c_{2x}^\dagger c_{2x}\right).
\end{align}
Now, CS still holds if in addition to $v(t)$ and $w(t)$ being even
functions of time, $u(t)$ is odd: $u(t)=-u(-t)$. We choose
$u(t)=\sin(2\pi t)$.  This time, if we break CS by shifting the
$v(t)$ pulse in time with respect to the $w(t)$ and $u(t)$ pulses, as
shown in Fig.~\ref{fig:break_cs}(b), all end states move away from
their original energies (again except for the time shift of $0.5$).

The extra PHS of the PDSSH model brings with it an extra pair of bulk
topological invariants, $(Q_0,Q_\pi) \in \mathbb{Z}_2 \times
\mathbb{Z}_2 $, which predict the number of end states protected by
PHS at $0$ and $\pi$ energy. If we have CS, the invariants are just
$Q_\varepsilon = \nu_\varepsilon \text{mod }2$; if CS is broken,
however, they can only be obtained by a procedure involving analytic
continuation based on the full cycle $H(t)$, as found by Jiang et
al. \cite{akhmerov_majorana_driven}.

We find that for the PDSSH model, the invariant of Jiang et
al. \cite{akhmerov_majorana_driven} can be given by simple closed
formulas.  At the momenta $k=0$ and $k=\pi$, the Hamiltonians at
different times all commute with each other, and therefore, all that
matters is the total area under the $v$ and $w$ pulses,
\begin{align}
V &= \int_0^1 v(t)dt;&   W &= \int_0^1 w(t) dt.
\end{align}
A short calculation gives  
\begin{align}
Q_0 &= \text{sgn} \left( \sin \frac{V+W}{2} \sin \frac{V-W}{2} \right);\\ 
Q_0 Q_\pi &= \text{sgn} \big( \sin (V+W) \sin (V-W) \big).
\label{eq:invariants_PHS}
\end{align}

\section{Mapping to the discrete time quantum walk}

The PDSSH model, besides being the simplest periodically driven
topological insulator, also gives a lattice realization of the
discrete time split-step quantum walk. For the quantum walk, we need
to define the basis states $\ket{R/L,x}$, for coin state predicting
the next step right/left, and the walker at position $x$. These basis
states are identified with states on the SSH chain as
\begin{align}
c_{2x+1}^\dagger \ket{0} &= \ket{R,x};\\
c_{2x}^\dagger \ket{0} &= -i \ket{L,x}. 
\label{eq:basis_qw}
\end{align} 
The basic operations of the split-step walk are rotations of the
internal state of the walker, $R(\theta)=e^{-i\theta \sigma_y}$, and
shifts of the $R/L$ internal state to the right/left, given by
$S_{\pm} = e^{-ik (\sigma_z \pm 1)}$.
One timestep of the split-step walk is defined as  
\begin{align}
U &= S_- e^{-i \theta_2 \sigma_y}
S_+ e^{-i \theta_1 \sigma_y}.
\end{align}
A pulse of $v$ of area $V$ followed by a pulse of $w$ of area $W$, in
the basis of Eq.~\ref{eq:basis_qw}, can be written as
\begin{align}
U &= e^{- i W (\cos k \sigma_y - \sin k \sigma_x)} e^{-i V \sigma_y},
\label{eq:recipe_walk_H}
\end{align}
which reproduces the timestep of the split-step walk with the angles 
\begin{align}
\theta_2 &= W+\pi/2;&  \theta_1 &= V-\pi/2.
\end{align}

The above mapping is important as it allows us to apply results about
the topological phases of periodically driven systems to quantum
walks. 

As an example, consider the invariants due to CS, via
Eqs.~\eqref{eq:asboth_tarasinski}, for the simple quantum walk, given
by $U=S_-S_+ e^{-i\theta \sigma_y}$. According to the mapping above,
the winding numbers are $\nu_0 = \nu[-i(s+ce^{ik}]$, $\nu_\pi =
\nu[c-se^{-ik}]$, with $c=\cos (\pi/4+\theta/2)$, $s=\sin
(\pi/4+\theta/2)$. We get $(\nu_0,\nu_\pi)=(+1,0)$ if
$\abs{c}>\abs{s}$, i.e., if $\theta\in [-\pi,0]$, and $(0,-1)$ if
$\theta \in [0,\pi]$.  
This is shifted by $(1/2, -1/2)$ from the invariants obtained by the
scattering matrix method\cite{scattering_walk2014}, but such a shift
is not physical: both methods predict a pair of end states at $0$ and
$\pi$ quasienergy at an interface between bulks with $\theta<0$ and
$\theta>0$, as seen in simulations \cite{asboth_prb}.

Another example is the calculation of the invariants due to PHS in the
split-step quantum walk. Compared to the invariants
$Q_0^{(\text{gap})},Q_\pi^{(\text{gap})}$, defined via gap closings in
the parameter space\cite{asboth_prb}, the above mapping to the PDSSH
model, together with Eqs.~\eqref{eq:invariants_PHS} gives
$Q_0=Q_0^{(\text{gap})}$, and $Q_\pi= 1-Q_\pi^{(\text{gap})}$, which
agrees in all the predictions concerning end states at
interfaces. Compared to the scattering matrix topological
invariants\cite{scattering_walk2014}, we of course find the same
constant shift by $(1/2,-1/2)$ as for the invariants due to CS, which
has no influence on the physical predictions.

\end{document}